\documentstyle[epsf,twocolumn,aps]{revtex}

\begin{document}
\draft
\preprint{HEP/123-qed}
\title{The Light Quark Masses from Lattice Gauge Theory
}
\author{ Brian J. Gough,$^{a,}$\cite{BG}
George M. Hockney,$^a$  Aida X. El-Khadra,$^b$ 
Andreas S. Kronfeld,$^a$ Paul B. Mackenzie,$^a$
Bart~P.~Mertens,$^c$ Tetsuya~Onogi,$^d$ and James N. Simone$^{a}$
}
\address{
$^a$ Fermilab, PO Box 500,  Batavia, Illinois 60510
}

\address{
$^b$ Dept. of Physics, University of Illinois, Urbana, Illinois  61801
}

\address{$^c$ Dept. of Physics, University of Chicago, Chicago, Illinois 60637
} 

\address{
$^d$ Dept. of Physics, University of Hiroshima, Higashi-Hiroshima 739, Japan
}

\date{\today}
\maketitle
\begin{abstract}

We investigate the masses of the light quarks with lattice QCD.
We show that most of the large  dependence on the lattice spacing $a$
 observed in previous  determinations  using Wilson
fermions is removed with the use of an  ${\cal O}(a)$ 
corrected action.
In the quenched approximation,
we obtain  for the strange quark $\overline{MS}$ mass
$\overline{m}_s(2\ {\rm GeV})	 = 	 95 (16)\ {\rm MeV}$,
and for the average of the up and down quark masses
$\overline{m}_l(2\ {\rm GeV})	 = 	 3.6 (6)\ {\rm MeV}$.
Short distance arguments and existing staggered fermion calculations make
it likely that the answers including the effects of quark loops lie
20\% to 40\% below this:
 $ \overline{m}_s(2\ {\rm GeV})$ in the range 54--92 MeV,
and 
$\overline{m}_l(2\ {\rm GeV})$ in the range 2.1--3.5 MeV.
We argue that almost all lattice determinations of the light quark masses
are  consistent with these   values.
These low values 
 are outside the range usually given by conventional phenomenology.
\end{abstract}



Among the most important applications of lattice gauge theory to particle 
physics
are the calculations required to determine the
fundamental parameters of the quark sector of the standard model.
One of the most important of these is the overall scale of the light quark
masses.
It is one of the least well known of the fundamental parameters of the standard
model.
(Estimates for the strange quark mass range  from
100 to 300 MeV
for the  $\overline{MS}$ masses renormalized at a ``high''
 energy scale,
1 GeV,
and for the average light quark mass from
3.5 to 11.5 MeV \cite{RPP94}.)
It is also one for which lattice methods are almost uniquely reliable,
unlike  quark mass ratios or the strong coupling constant $\alpha_s$,
for which other powerful methods exist.
Values for quark masses have been obtained since almost the beginning
of lattice phenomenology \cite{Uka93,Gup94}.
However, improved understanding of perturbation theory and finite lattice
spacing errors has been required to make sense of the various
lattice determinations,
which initially ranged over a factor of three.

Lattice determinations of standard model parameters consist of two pieces.
Calculations of experimentally measurable  quantities such as hadron
masses are used to fix the bare coupling constant and quark masses
 in the lattice Lagrangian.
Short distance calculations are used to relate the bare 
 parameters in the lattice theory to renormalized, running
coupling constants and masses, such as those of
the $\overline{MS}$ scheme.

Quark masses are most easily obtained in lattice calculations by matching
pseudoscalar meson masses with experiment.
These are among the easiest lattice calculations,
having  small statistical and finite volume errors.
Experimental uncertainties are  also negligible.
Uncertainties are dominated by truncation of perturbation theory
 and discretization errors,
 and by errors arising from
the omission of light quark loops (the ``quenched'' approximation).

The short distance calculations relating the parameters in various 
regulators may be performed by demanding that short distance quantities
such as the heavy-quark potential or current correlation functions
be the same in both regulators.
It is desirable to do the lattice part of such calculations nonperturbatively
as much as possible, to test for the presence of nonperturbative 
short distance effects and possible poor convergence of perturbation theory.
Such nonperturbative short distance analysis for quark masses
is  currently less advanced than the analogous investigations 
for the strong coupling constant.

Perturbative relations between the lattice bare mass, $m_0$,  and the 
$\overline{MS}$ mass, $\overline{m}$, 
may be obtained by demanding that on-shell
Green's functions calculated with both regulators be equal.
Analogous perturbative expressions for the renormalization
of $\alpha_s$ were initially rendered almost useless by sick behavior in 
the lattice perturbation series.
In Ref. \cite{LM93} it was shown that such behavior could be understood
and mostly eliminated by a mean field theory resummation of large
``tadpole'' graphs.
The effects of such large tadpoles are much less important for quark 
mass renormalizations than for $\alpha_s$ \cite{endnote}.

To  reduce the effects of such graphs further, 
the expression giving  $\overline{m}$ from  $m_0$
may be rewritten in terms of a
 mean field improved mass $\tilde{m}$,
\begin{eqnarray}\label{m0msbar}
\overline{m}(\mu)
&=& \tilde{m}\left[1+\alpha_s  \gamma_0  \left(  \ln \tilde{C}_m-\ln\left( a \mu\right) \right) \right],  \label{eq:star}
\end{eqnarray}
where $\gamma_0=2/\pi$ is the leading quark mass
anomalous dimension, and $\ln \tilde{C}_m$ is the result of 
a one loop calculation.
Here we use $\tilde{m}=m_0/\sqrt[4]{\langle U_P\rangle}$
for the  mean-field-improved bare mass $\tilde{m}$.
The nonperturbative value of the plaquette expectation value 
$\langle U_P\rangle$ 
is used in the expression for $\tilde{m}$ to incorporate an estimate
for higher order tadpole graphs.
 The one loop term $\ln \tilde{C}_m$ is then adjusted to remove the one loop
part of this expression 
$\sqrt[4]{\langle U_P\rangle}=(1-(\pi/3)\alpha_s)$.

\begin{figure}
\epsfxsize=0.45 \textwidth
\epsfbox{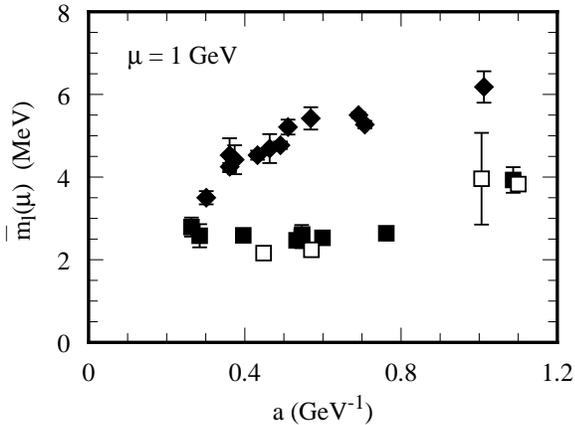}
\caption[oldmasses]{Previous lattice results for the $\overline{MS}$ masses of 
the light quarks, renormalized at 1 GeV, with 
the lattice spacing set by the rho mass.
  Lattice spacing dependence is large for
quenched Wilson fermions (diamonds) and small  for
quenched staggered fermions (filled squares).
 Results from two-flavor staggered fermion QCD (open squares)
lie below those from   quenched approximation staggered fermions
by a reasonable amount.
Data from Ukawa \cite{Uka93}.
}
\label{oldmasses}
\end{figure}

In Fig. \ref{oldmasses} we show a compilation of previous results 
 given  by Ukawa
\cite{Uka93}.
Quenched results obtained with staggered fermions 
 are almost cut-off independent for lattice
spacings less than 1 GeV$^{-1}$.
However, for staggered fermions the constant
   in Eq. (\ref{m0msbar}) is
$\tilde{C}_m= 132.9$ \cite{Gro84}.  
This leads to correction factors of 50--100\%, most of which is
unexplained by mean field theory,
casting doubt on the reliability of the perturbative relation between
the staggered-fermion quark mass and the $\overline{MS}$ quark mass.

For Wilson fermions, we have    $\tilde{C}_m = 1.67$ \cite{Gro84},
and thus a well-behaved perturbation series.
However, the numerical results for the Wilson action
 show large cut-off dependence.
They lie far above the results for staggered fermions
 but show a downward trend as the lattice spacing is reduced.
The Wilson fermion action contains an error of ${\cal O}(a)$, which is
absent in the staggered fermion action.
If the results are extrapolated in $a$, one obtains
 a result much closer to the results of staggered fermions.
(See, for example, Ref. \cite{Gup96}.)
However,  remaining sources of cut-off dependence are an unknown combination
  of ${\cal O}(\alpha_s^2)$, 
${\cal O}(\alpha_s a)$, ${\cal O}(a^2)$, etc.  They  cannot be 
estimated or removed
by simple extrapolation,
since we do not have a quantitative theory of their functional form.
One therefore needs to investigate the removal of the dominant 
${\cal O}(a)$ error  from the Wilson action.

A convenient action for doing this
has been proposed by Sheikholeslami and Wohlert~\cite{She85}.
Their improved action incorporates an extra dimension five term
$ \overline{\psi} \sigma_{\mu\nu} F_{\mu\nu} \psi$,
the so-called ``clover'' term.
The one-loop correction to the coefficient of the clover term
 is large \cite{lue96},
as suggested by  mean field theory \cite{LM93}.
It is a three-tadpole correction and can be approximated by
$c \approx \langle U_P\rangle^{-3/4},$ where the tree level coefficient is 
normalized to be one.
For the improved action, $\tilde{C}_m=4.72$  \cite{Gab91}.
Thus, Eq.~(\ref{m0msbar}) is still well--behaved.

We  use this action to determine the overall scale of the light
quark masses.
(Or equivalently, the coefficient of $m_l$ in the expression
$M_\pi^2 = C m_l + \ldots$.  
We do not see deviations from the leading order of this equation,
see below.)
Our  lattice spacings range from (the coarse) 1.26  GeV$^{-1}$
(at which uncertainties due to perturbation theory are starting to approach
50\%), down to 0.39 GeV$^{-1}$ (where perturbation theory appears well
behaved).
We have performed the calculation
 at the largest lattice spacing to investigate its
 behavior where it is beginning to break down,
but we omit it from our final results.
The lattice spacings have been obtained from the 1P-1S splitting of the
charmonium system, $M_{h_c} - (3 M_{J/\psi} + M_{\eta_c})/4  $,
for which the uncertainties of lattice calculations are particularly
small and easy to understand.
This means that numerical uncertainties in our results for the quark masses
arise from a combination of uncertainties in the charmonium and pion
calculations.

We  use improved lattice perturbation theory to convert to the
$\overline{MS}$ mass at renormalization scale $\mu=2$ GeV and charmonium
splittings to determine the lattice spacing, 
whereas previous
determinations  typically used bare perturbation theory
at  scale $\mu = 1$
 GeV and the rho meson mass to determine the lattice spacing.
Although renormalization at 1 GeV is conventional in nonlattice 
results, renormalizing down to such a low scale
 introduces additional perturbative uncertainty into
the results which is not present in the underlying lattice results.

Discussions of our charmonium calculations have appeared in 
Ref. \cite{El-Khadra+al}.
Some technical details and results of our pion calculations are 
given in Table \ref{tab:2}.
For our most significant data point, the improved clover action at
$\beta=6.1$, we have used 100 configurations separated by 4000 
heat bath gauge sweeps.  Pion correlation functions were calculated
using $2\times 2$ correlated fits (fitting two states using two operators
for the pions).  Contributions from excited states were checked further
on the smaller lattices by comparing  with $1\times 1$ and $3\times 3$ fits.
Statistical errors were calculated using 1000 bootstrap samples.
Longer descriptions of our analyses for the charmonium system
 and for the light quark masses are in preparation~\cite{big}.

\begin{figure}[t]
\epsfxsize=0.45 \textwidth
\epsfbox{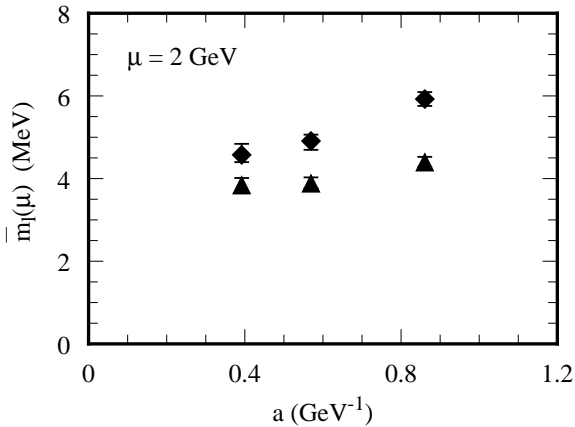}
\caption[masses]{Our results for the masses of 
the light quarks.
 Most of the lattice spacing dependence of 
unimproved Wilson fermions (diamonds)
is removed by the use of an ${\cal O}(a)$ corrected action 
(triangles)
with a tadpole improved coefficient.
The lattice spacing is set by the charmonium 1P-1S splitting.
Errors are statistical only.
}
\label{masses}
\end{figure}

In Fig. \ref{masses} and in Table \ref{tab:2} we show our 
results for the light quark masses in the quenched approximation.
The errors shown are statistical only.
The diamonds  are our results for unimproved Wilson fermions.
They are consistent  with the existing determinations (diamonds
 in Fig.~\ref{oldmasses}).
The triangles are our results for the mean-field-improved
clover action.
 Most of the cut-off dependence has been removed.
 
Remaining sources of such cut-off dependence could include
large $\alpha_s^2$ corrections to the mass relation, Eq. (\ref{eq:star}),
further corrections to the clover coefficient in
the pion numerical calculations, and ${\cal O}(a^2)$ corrections to
the charmonium 1P-1S splitting. ${\cal O}(a)$ corrections are expected to
be negligible for this splitting, but ${\cal O}(a^2 p^2)$ corrections
could be larger since quark momenta are 
 larger in charmonium than in pions.
We estimate ${\cal O}(a^2 p^2)$ corrections in the charmonium splitting
to range from 4\% to 20\% on our three finest lattice spacings.
The perturbative one-loop result for the coefficient of the 
${\cal O}(a)$ clover correction
agrees with the mean field estimate \cite{lue96}.
However,
 a  nonperturbative determination 
appears indeed to favor a further significant correction~\cite{Jan96}.
Purely perturbative errors in the relation between the lattice and
$\overline{MS}$ masses
should be of order $\alpha_s^2 \sim 5\%$
at our finest lattice spacing.
Other smaller uncertainties include finite volume effects, which are expected
to be a couple of per cent or less, and statistical errors, 
which are 4\% and arise mostly from the lattice spacing derived from the
charmonium system.

We have examined the pseudoscalar meson mass squared as
a function of the quark mass, which should be linear plus small
corrections in the small quark mass limit.
Our numerical data are for quark masses in the range $0.4 m_s$ to $m_s$.
In this range, we find no statistically significant evidence for
 quadratic terms in $M_\pi^2$ vs. $m_l$, much less the very large
quadratic terms that have been
postulated to make $m_u=0$ consistent with experiment.
Therefore, our results for the ratio of the strange to light quark masses
agree with lowest order chiral perturbation theory:
$(m_s+m_l)/(2m_l)\approx M_{K^0}^2/M_{\pi^0}^2\approx 13.6$.
The values for $m_l$ are obtained by linear extrapolation from the 
lowest masses at which we have  performed simulations
down to the physical mass value.

At present,
the uncertainty associated with the remaining cut-off dependence is 
the least reliably understood uncertainty in the quenched 
approximation.
Pending further understanding of this error,
we take our result at the smallest lattice spacing as 
the top of our lattice spacing error bar.
We take a linearly extrapolated result (the lower of two  plausible
extrapolation methods) through the three finest lattice spacings as the bottom.
This gives a range of 0.8 MeV for the cut-off dependence uncertainty, 
and we take the center of this range as our
continuum limit, quenched approximation result:
\begin{eqnarray}\label{eq:qu}
\overline{m}_l(2\ {\rm GeV})	& = 	& 3.6 (6)\ {\rm MeV},\\
\overline{m}_s(2\ {\rm GeV})	& = 	& 95 (16)\ {\rm MeV}.
\end{eqnarray}
The perturbative and cut-off dependence uncertainties were added linearly
in the total error,
since they are related.
All other uncertainties were added in quadrature. 

Another determination of the strange quark mass with an ${\cal O}(a)$
improved  action has been reported \cite{All94}.
This determination used a tree-level, rather than a mean-field improved,
estimate for the clover coefficient.
They obtained $\overline{m}_s(2\ {\rm GeV})=128 (18)$ MeV.
They did not attempt to
 correct for the effects of the remaining lattice  spacing dependence
or the effects of the quenched approximation.
Most of the discrepancy with our results arises from fact that we have used
much larger clover coefficients, and make an allowance for the fact that 
we continue to find significant cut-off dependence even so.

In the quenched approximation, QCD couplings run slightly incorrectly.
The strong coupling constant, for example, runs too fast
without the effects of light quark loops~\cite{El-Khadra+al}.
To leading logarithmic accuracy, $\alpha_s(\pi/a)$ is too small by a factor
of about $\beta_0^{(3)}/\beta_0^{(0)}$,
where $ \beta_0^{(0)}$ and $\beta_0^{(3)}$ are 
 the leading quenched and unquenched $\beta$ functions,
respectively.
This means that the running of the quark mass in the perturbative momentum
region around $\pi/a$ is too slow, by about the same factor.
In Ref. \cite{Mac94}, the ratio of quenched and unquenched quark masses arising
from the perturbative region was estimated, to leading logarithmic accuracy,
to be
\begin{eqnarray}
\frac{m(\pi/a)|_{\rm qu.\ \ \ }}{m(\pi/a)|_{\rm unqu.}}
&\approx& \alpha_s(\pi/a)^{\frac{\gamma_0}{2}(1/\beta_0^{(0)}-1/\beta_0^{(3)})}\\
&\approx& 1.15 {\ \rm to \ } 1.20,  \label{eq:1.15}
\end{eqnarray}
for $\alpha_s(\pi/a)\approx 1/6$ to 1/8.
There is, of course, an additional contribution from the nonperturbative 
region, which is unknown.
However, a correction due to light quark loops of tens of per cent in
the downward direction
from the perturbative region at least would not be unexpected.

Some quenched and unquenched staggered results summarized in Ref. \cite{Uka93} 
are
shown in Fig.~\ref{oldmasses}.
(Unquenched Wilson fermion calculations appear to be much more difficult
to perform and harder to interpret.)
The unquenched results indeed lie below the quenched results by roughly 
the expected amount, and we  take them seriously enough to use them
to estimate the effects of  quenching.
We argued above that quenched staggered quark mass determinations look
 good in most ways, but are unreliable because of the
 poor convergence of perturbation theory.
However, the large corrections cancel out in the ratio of the quenched
and unquenched determinations, making this a useful quantity to examine.
To minimize effects due to differences in analysis methods, we estimate
the ratio from the results of a single group, at similar volumes and
lattice spacings (about 0.4 GeV$^{-1}$)  \cite{Ish92,Fuk92}, and obtain
\begin{eqnarray}
\frac{\overline{m}_l(1.0\ {\rm GeV})_{\rm n_f=0}}{\overline{m}_l(1.0\ {\rm GeV})_{\rm n_f=2}}
&\approx& \frac{2.61(9)}{2.16(10)}\\
&=& 1.21(7)  \label{eq:unq}.
\end{eqnarray}
Since there are, in fact, three flavors of light quarks in the world 
and not two, we will use this ratio as a lower bound on the actual ratio
and use the square (corresponding to four light quarks) as an upper bound.

In summary, after making some plausible cuts,
existing determinations are reasonably
consistent, or have plausible explanations for discrepancies.
We omit results with very small physical volumes
(smaller than 1.5 fm) and very large lattice spacings 
(larger than 0.2 fm, or 1.0 ${\rm GeV}^{-1}$).
We also do not attempt to interpret the results 
 with unquenched Wilson fermions, which are in a more primitive state
than those with staggered fermions.
Of the remaining determinations, we have shown that the  cut-off
dependence and large size of determinations with quenched Wilson
 fermions arise mostly from the well-known ${\cal O}(a)$ error.
The remaining  discrepancy between the quenched clover-improved fermion results
and the quenched staggered fermion results is plausibly attributed to the
apparent poor convergence of staggered fermion perturbation theory
and the remaining cut-off dependence in the improved fermion results.
The small difference between quenched and unquenched staggered 
fermion results
 is roughly what is expected.
Putting all this together, we  arrive at the following estimates
for the light quark masses including effects of light quark loops, 
which we believe are consistent with
all known facts:
\begin{itemize}
\item $\overline{m}_s(2\ {\rm GeV})$ in the range  54--92  MeV, 
\item $\overline{m}_l(2\ {\rm GeV})$ in the range  2.1--3.5 MeV,
\end{itemize}
for the $\overline{MS}$ masses renormalized at 2 GeV.
These estimates arise from combining our quenched result, Eq.~(\ref{eq:qu}),
with the correction ratio obtained from staggered fermions, 
Eq. (\ref{eq:unq}).
Renormalizing down to the scale 1 GeV, where conventional mass estimates are
often quoted, the estimates are raised by 10\%, to 
 $\overline{m}_s(1\ {\rm GeV})$ in the range 59--101~MeV, and 
$\overline{m}_l(1\ {\rm GeV})$ in the range  2.3--3.9~MeV.

\acknowledgments
We  thank Akira Ukawa for sharing the data in Ref. \cite{Uka93}
with us.
We  thank him,  Steve Gottlieb, and Peter Lepage
for helpful conversations.
High-performance computing was carried out on ACPMAPS, which is
operated and maintained by the High Performance and Parallel Computing
and Electronic Systems Engineering Departments of Fermilab's Computing
Division; we thank past and present members of these groups for making
this work possible.
AXK was supported in part by the DOE OJI program under contract
no. DE-FG02-91ER40677.
TO would like to thank the Nishina Foundation
  for  support during his visit to Fermilab.
Fermilab is operated by Universities Research
Association, Inc. under contract with the U.S. Department of
Energy.

\begin{table}[hbt]
\caption{Our results for $\overline{m}_l(2\ {\rm GeV})$, the average of the $u$ and $d$ quark masses,
renormalized at 2 GeV. 
$c$ is the coefficient of the ${\cal O } (a)$ improvement operator.}
\label{tab:2}
\begin{tabular}{l|cccc}
$\beta$	&	5.5 & 	5.7	&	5.9&	6.1 \\
\hline
$a$\ (GeV$^{-1}$) &1.26	& 0.86	& 0.57	& 0.39  \\
volume & 8$^3\times 16$	& 12$^3\times 24$& 16$^3\times 32$& 24$^3\times 48$  \\
$m$ ($c=0$) & 6.31(26)	& 5.93(17)	& 4.88(18)	& 4.62(22)  \\
$m$ (improved) & 4.75 (19) & 4.41(12) & 3.90(13)	&3.84(18)  \\
$c$  & 1.69 	& 1.57	& 1.50	& 1.40  \\
\end{tabular}
\end{table}

\end{document}